\documentclass[nofootinbib,superscriptaddress]{revtex4}
\usepackage{amsmath,amsfonts,amssymb,color,adjustbox,lipsum,slashed,subfig,subfloat,graphicx}
\definecolor{blu}{cmyk}{1,0.7,0,0.6}
\begin{document}
\title{\color{blu}Probing radiative neutrino mass models with dilepton events at the LHC}
\author{Chahrazed Guella}
\email{chahra.guella@gmail.com}
\affiliation{Facult\'e de Physique, d\'{e}partement de G\'enie Physique, USTO-MB, BP1505 El
M'Naouar, Oran, Alg\'erie.}
\author{Dounia Cherigui}
\email{dounia.ch13@gmail.com}
\affiliation{Facult\'e de Physique, d\'{e}partement de G\'enie Physique, USTO-MB, BP1505 El
M'Naouar, Oran, Alg\'erie.}
\author{Amine Ahriche}
\email{aahriche@ictp.it}
\affiliation{Laboratory of Mathematical and Sub-Atomic Physics (LPMPS), University of
Constantine I, DZ-25000 Constantine, Algeria}
\affiliation{The Abdus Salam International Centre for Theoretical Physics, Strada Costiera
11, I-34014, Trieste, Italy.}
\affiliation{Department of Physics and Center for Theoretical Sciences, National Taiwan
University, Taipei 106, Taiwan.}
\author{Salah Nasri}
\email{snasri@uaeu.ac.ae}
\affiliation{Department of Physics, United Arab Emirates University, Al-Ain, UAE.}
\author{Rachik Soualah}
\email{rsoualah@cern.ch}
\affiliation{Department of Applied Physics and Astronomy, University of Sharjah, P. O. Box
27272 Sharjah, UAE.}
\affiliation{The Abdus Salam International Centre for Theoretical Physics, Strada Costiera
11, I-34014, Trieste, Italy.}
\affiliation{INFN, Sezione di Trieste, Gruppo collegato di Udine, Italy.}

\begin{abstract}
\vspace{1.5cm}

In this work, we investigate the possibility of probing a class of neutrino
mass models at the LHC proton-proton collisions with 8 and 14 TeV energies.
The existence of lepton flavor violating interactions for a singlet charged
scalar, $S^{\pm}$, that couples to the leptons could induce many processes
such as $pp\rightarrow\ell_{\alpha}^{\pm}\ell_{\beta}^{\mp}+\slashed E $.
Using the processes with $\ell_{\alpha}\ell_{\beta}=ee,e\mu,\mu\mu$, we found
that an inclusive cut on the $M_{T2}$ event variable is vital in our analysis
and leads to an effective suppression of the large Standard Model background.
Our results show possible detectability of the charged scalars effect,
especially at the $\sqrt{s}= 14~\text{TeV}$.

\vspace{1cm} \textbf{PACS}: 12.60.-i, 12.60.Jv, 14.80.Cp, 12.15.-y.\newline%
\vspace{0.5cm} \textbf{Keywords}: charged scalar, missing energy,
significance. \vspace{0.5cm}

\end{abstract}
\maketitle

\section{Introduction}

The Standard Model (SM) has been very successful in describing nature at the
weak scale and it is in very good agreement with most of the experimental
results. However depending on some of astrophysical and cosmological
observations, many questions remain unanswered such as the nature of the
nonbaryonic matter (dark matter), neutrino masses and their mixing, hierarchy
problem, and matter-antimatter asymmetry of the Universe. Therefore, these
limitations lead to the need to introduce additional components to the SM
and go beyond its actual framework. There is no unique way to introduce
neutrino masses in the SM. One of the attractive mechanisms that can explain
the smallness of neutrino masses is the radiative mass generation mechanism.
In this way, neutrinos are massless at the tree level and get a nonzero
mass at the loop level which can be naturally small due to the extra loop
suppression factors. This can be implemented at one loop~\cite{Zee-Ma},
two loops \cite{Zee-86}, or three loops~\cite{Aoki:2008av, knt}, \footnote{The
Krauss-Nasri-Trodden model~\cite{knt} has been generalized where the
singlets $\mathcal{Z}_{2}$ odd singlet charged $S_{2}$, and the RH Majorana
neutrinos $N_{i}$ are promoted to triplets \cite{knt3}, quintuplet~\cite{knt5}%
, septuplets~\cite{knt7}, and in a scale-invariant
framework~\cite{Ahriche:2015loa}. These models phenomenology have been
investigated in Refs. ~\cite{hna},~\cite{ILC} and \cite{AMN}.} which can be achieved
by extending the SM with new interactions involving additional scalar singlets
and/or doublets.

In Ref.~\cite{hna}, the SM was extended with two electrically charged singlet
fields under $SU(2)_{L}$ scalars and three right-handed (RH) neutrinos,
$N_{i}$ where a $Z_{2}$ symmetry was imposed to forbid the Dirac neutrino mass
terms at tree level \cite{knt}. Once the electroweak symmetry is broken,
neutrino masses are generated at three loops, and the lightest RH neutrino,
$N_{1}$, could be a candidate for dark matter (DM). Some generalizations of
this model have been proposed where one of the two singlet charged scalars and
$N_{i}$ are replaced by $SU(2)_{L}$ triplet in Ref. \cite{knt3} or by $SU (2)_{L}$
quintuplet fields~\cite{knt5}. The scalar spectrum of this model has a pair of
charged scalar particles which can be produced at colliders via their
couplings to photons and the SM higgs boson. The signatures of charged scalars
at LHC and the future high-energy colliders have
been studied by many authors (see, for example Refs.~\cite{osland, Cheung, FLV-bound,
seesaw}).

The main goal of our work is to present the production mechanism of the
singlet charged scalar. Moreover, we show that by maximizing the event number
excess of the dilepton process $pp\rightarrow\ell_{\alpha}^{+}\ell_{\beta}%
^{-}+\slashed E$ at the LHC could be a direct probe of the charged scalar. We
will confront the model parameter space with the neutrino oscillation data and
the recent experimental bounds on the lepton flavor violating (LFV) processes,
such as the upper limit on the $\mu$ $\rightarrow$ $e$$\gamma$ provided by the
MEG Collaboration~\cite{Adam:2013mnn}. Furthermore, we impose appropriate
kinematic cuts on the outgoing leptons to optimize the signal in our model
over the SM background at the LHC Run I and Run II energies.

The rest of the paper is structured as follows. In Sec II ,
we present the class of models for neutrino mass where we describe
the decay and production of the charged scalar at the LHC for c.m. (CM) energies $\sqrt {s}=8~\text{\textrm{TeV}}$ and
$\sqrt{s}=14~\text{\textrm{TeV}}$. Section III is devoted to the
study of signatures of the charged scalars at the LHC and the
significance of the signal as a function of the expected luminosity.
Finally, we give our conclusion in Sec. IV.

\section{Charged Scalars in Radiative Neutrino Mass Models}

To generate neutrino mass at the loop level, the SM is extended by extra
scalars and fermions, among which is a charged scalar, $S^{\pm}$, that
transforms under the gauge group $SU(3)\times SU(2)_{L}\times U(1)_{Y}$ as
$S^{+}\sim(1,1,2)$ with the following interactions in the Lagrangian
\begin{align}
\mathcal{L}  &  \supset\{f_{\alpha\beta}\overline{L_{\alpha}^{c}}L_{\beta
}S^{+}+\mathrm{H.c.}\}-M_{S}^{2}S^{+}S^{-}-V(H,S,\phi_{i}),\label{L}\\
V(H,S,\phi_{i})  &  \supset\lambda_{HS}\left\vert H\right\vert ^{2}\left\vert
S\right\vert ^{2}%
\end{align}
where $L_{\alpha}=(\nu_{\alpha L},\ell_{\alpha L})^{T}$, $\ell_{\alpha R}$ is
the charged lepton singlet, $f_{\alpha\beta}$ are the Yukawa couplings which
are antisymmetric in the generation indices $\alpha$ and $\beta$, and $c$
denotes the charge conjugation operation. Here, $H$ represents the SM Higgs
field doublet, $V(H,S,\phi_{i})$ is the scalar potential and $\phi_{i}$\ could
be any additional scalar representation(s) that is (are) required to generate
the neutrino mass at loop level.

In such a model, the parameter space must accommodate neutrino masses and mixing
and at the same time satisfy the LFV constraints. Also, the charged scalar
fields contribute to the Higgs decay channel $h\rightarrow\gamma\gamma$ and
might lead to an enhancement with respect to the SM, whereas $h\rightarrow
\gamma Z$ is reduced. Furthermore, a strong first-order electroweak phase
transition can be achieved through the coupling of the new scalar degrees of
freedom (i.e., $S$ and $\phi_{i}$) around the electroweak scale to the SM Higgs
without being in conflict with the recent Higgs mass measurements provided by
the ATLAS~\cite{ATLAS} and CM~\cite{CMS} collaborations. In some models, when
a global symmetry $\mathcal{Z}_{2}$ is imposed, there exists a neutral
Majorana particle that can be a DM candidate. The collider phenomenology of
such kind of models is very rich and, in principle, can be probed through
various signals.

The new interactions in Eq. (\ref{L}) induce LFV effects via processes such as
$\mu\rightarrow e\gamma$ and $\tau\rightarrow\mu\gamma$ and a new contribution
to the muon's anomalous magnetic moment:%
\begin{align}
Br (\mu &  \rightarrow e\gamma)\simeq\frac{\alpha\upsilon^{4}}{384\pi}%
\frac{|f_{\tau e}^{\ast}f_{\mu\tau}|^{2}}{M_{S}^{4}},\label{meg}\\
Br (\tau &  \rightarrow\mu\gamma)\simeq\frac{\alpha\upsilon^{4}}{384\pi}%
\frac{|f_{\tau e}^{\ast}f_{\mu e}|^{2}}{M_{S}^{4}},\label{tmug}\\
\delta a_{\mu}  &  \sim\frac{m_{\mu}^{2}}{96\pi^{2}}\frac{|f_{e\mu}%
|^{2}+|f_{\mu\tau}|^{2}}{M_{S}^{2}}, \label{da}%
\end{align}
where $\alpha=e^{2}/4\pi$ is the fine structure constant, and $\upsilon
=246~\mathrm{GeV}$ is the vacuum expectation value of the neutral component in
the SM scalar doublet. The LFV branching ratios should not exceed the
experimental upper bounds $\mathcal{B}\left(  \mu\rightarrow e+\gamma\right)
<5.7\times10^{-13}$~\cite{Adam:2013mnn} and $\mathcal{B}\left(  \tau
\rightarrow\mu+\gamma\right)  <4.8\times10^{-8}$~\cite{PDG}. According to the
LFV experimental constraints, the couplings $f$ must scale like
$f\lesssim\varsigma M_{S}$, with $\varsigma$ as a dimensionful constant that
depends on each bound. This means that the couplings $f$ are suppressed for
low charged scalar mass values.

Here, we shall discuss the most striking features of the production mechanism
and decay modes of the singlet charged scalars that can be manifested at the
Large Hadron Collider. This extends our earlier studies~\cite{ILC} at the
future $e^{-}e^{+}$ colliders such as the International Linear Collider by including more accurate signatures for any possible charged scalars
production\footnote{A similar study has been performed in Ref. \cite{kanemura}.}. In
our analysis, we take $M_{S}$ to be within the range $[100\text{ \textrm{GeV}%
}-3\text{ \textrm{TeV}}]$ and select the parameter space to be in agreement
with the LFV bounds \cite{Adam:2013mnn}.

\begin{figure}[h]
\begin{centering}
\includegraphics[width = 11cm]{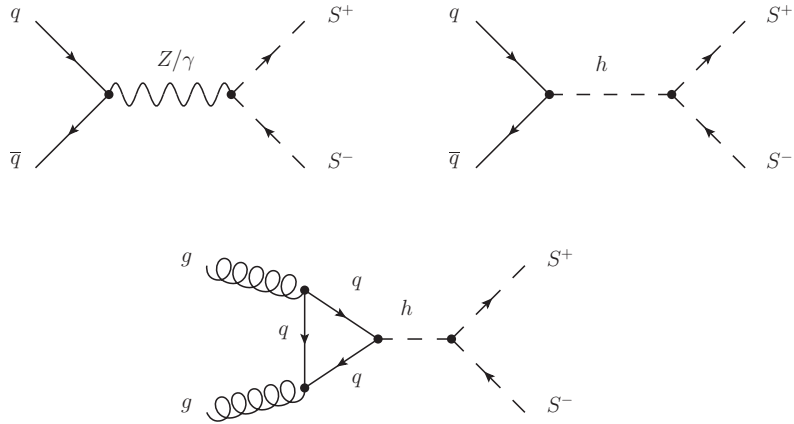}
\par\end{centering}
\caption{The Feynman diagrams of the charged scalar $S^{\pm}$ pair production
via the gluon-gluon fusion and quark-antiquark annihilation at proton-proton
collisions.}%
\label{prod}%
\end{figure}

In proton-proton collisions, the charged scalars $S^{\pm}$ can be produced in
pairs through the processes:
\begin{equation}
q\overline{q}\rightarrow\gamma/Z/h\rightarrow S^{+}S^{-},~gg\rightarrow
h\rightarrow S^{+}S^{-},
\end{equation}
which are depicted in Fig.~\ref{prod}. However, the contribution of the
quark-antiquark Higgs-mediated annihilation process is highly suppressed by
the small Yukawa coupling of the light quarks. Thus, the production is
expected to be dominated by the Drell-Yan process with the high-energy
partonic cross section given by\footnote{Since we are considering the charged
scalar masses much heavier than $M_{Z}$, one can safely neglect the Feynman
diagram with the Z gauge boson propagator.}
\begin{equation}
\hspace{-0.8cm}\sigma^{(DY)}(\hat s)=\frac{\pi\alpha^{2}}{9\hat{s}}\beta
^{3}\left[  e_{q}^{2}+\left(  \frac{2e_{q}g_{V_{q}}}{\cos^{2}{\theta_{\omega}%
}}\right)  \frac{\hat{s}}{(\hat{s}-M_{Z}^{2})}+\left(  \frac{g_{V_{q}}%
^{2}+g_{A_{q}}^{2}}{\cos^{4}{\theta_{\omega}}}\right)  \frac{\hat{s}^{2}%
}{(\hat{s}-M_{Z}^{2})^{2}}\right]  ,
\end{equation}
where $\hat{s}$ is the center-of-mass energy square of the parton system,
$\beta=\sqrt{1-4M_{S}^{2}/{\hat{s}^{2}}}$, $e_{q}$ is the electric charge
of the quark, ${\theta_{\omega}}$ is the weak mixing angle, and $g_{V}$ and
the $g_{A}$ are the SM vector and axial neutral current couplings.

\begin{figure}[h]
\begin{centering}
\includegraphics[width=0.5\textwidth]{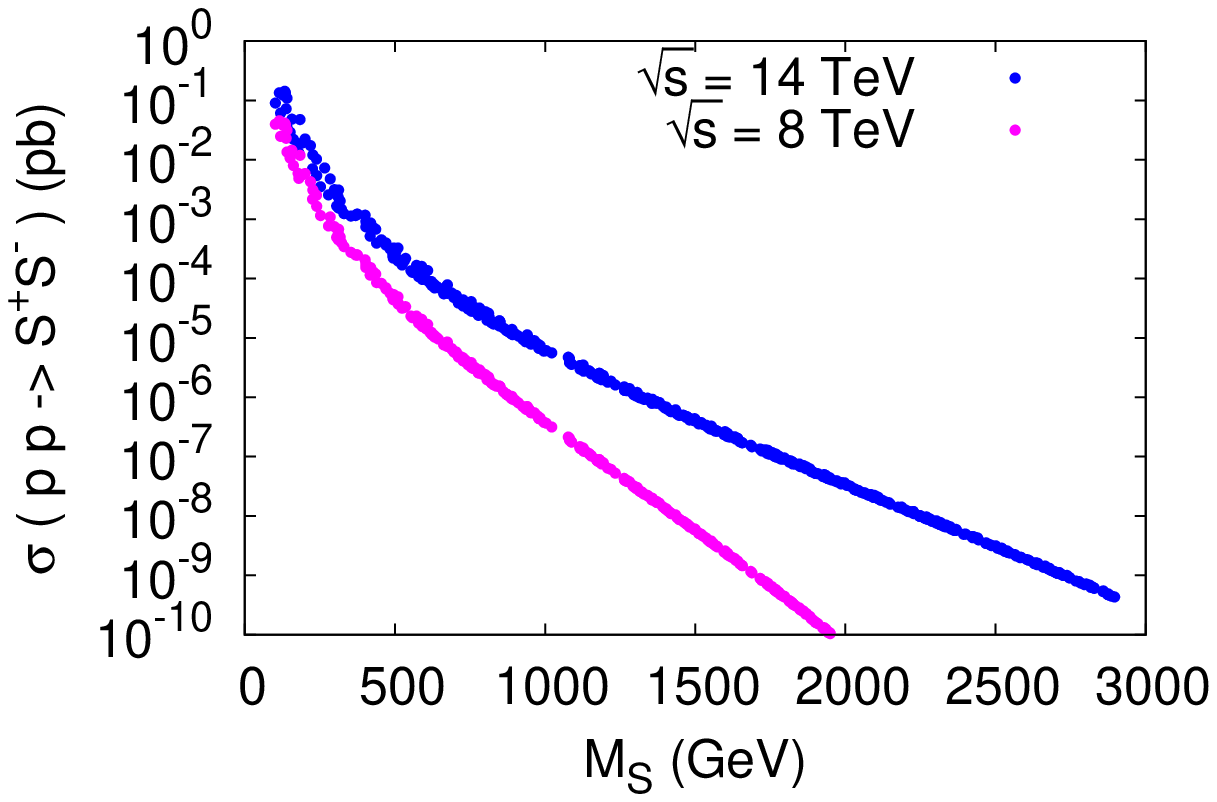}~\includegraphics[width = 0.5\textwidth]{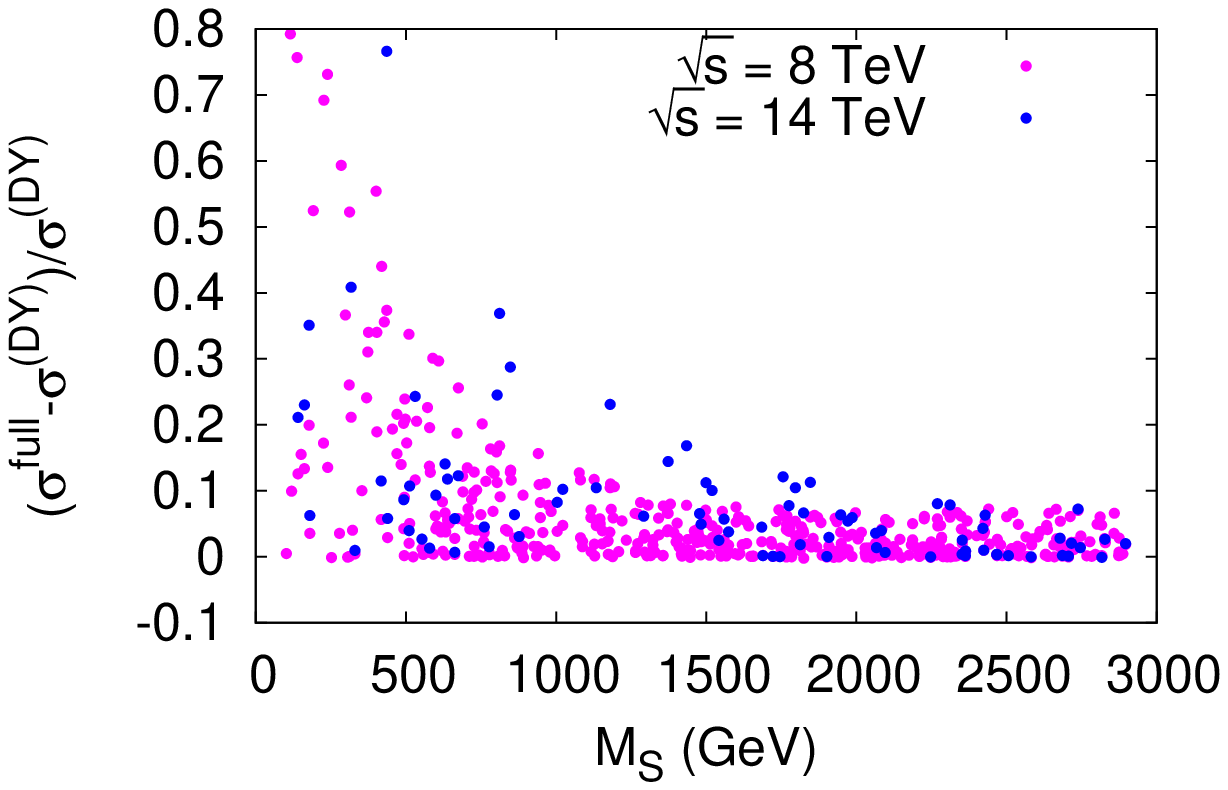}
\par\end{centering}
\caption{Left: the production cross section (in fb units) of $S^{+}S^{-}$ pair
production at $\sqrt{s}$ = 8 TeV (in magenta) and 14 TeV (in blue) vs the
charged scalar mass $M_{S}$. The dashed black lines represent the DY values.
Right: the ratio $\left[  \sigma^{full}(s)-\sigma^{(DY)}(s)\right]
/\sigma^{(DY)}(s)$ vs the charged scalar mass ($M_{S}$). The considered
$M_{S}$ values are in agreement with different experimental constraints such
as LFV and the Higgs decay $h\rightarrow\gamma\gamma$. }%
\label{knt-p}%
\end{figure}

For a random scan of the model parameter ($f_{\alpha\beta}$ and $M_{S}$),
taking into account the LFV constraints and the bounds on the Higgs decay
$h\rightarrow\gamma\gamma$, we show in Fig.~\ref{knt-p}, left, the cross section
of the singlet charged scalars pair production as a function of the charged
scalar mass ($M_{S}$) at both $\sqrt{s}=8$ and $14~\mathrm{TeV}$, where the
couplings $\lambda_{HS}$ are taken of order $\mathcal{O}(1)$. To
probe how dominant the DY contribution is, we show in Fig. \ref{knt-p}, right,
the ratio $\left[  \sigma^{(Full)}(s)-\sigma^{(DY)}(s)\right]  /\sigma
^{(DY)}(s)$ that characterizes the presence of the Higgs-mediated Feynman
diagrams, where $\sigma^{(Full)}(s)\equiv\sigma(pp\rightarrow S^{+}S^{-})$
including the Higgs exchange diagrams.

One notices that the production cross section depends very strongly on the c.m.
energy where it is more than 2 orders of magnitude larger for $\sqrt
{s}=14~{\text{TeV}}$ than at the $8~{\text{TeV}}$ c.m. energy. From Fig.
\ref{knt-p}, right, we see that the DY contribution may not be the dominant one,
especially for charged scalars lighter than 500 GeV.

After being produced, the charged scalar decays into a light neutrino and a
charged lepton with the partial decay rate as
\begin{equation}
\Gamma(S^{\pm}\rightarrow\nu_{\alpha}\ell_{\beta}^{\pm})=\frac{|f_{\alpha
\beta}|^{2}}{4\pi}~M_{S}\left(  1-\frac{m_{\ell_{\beta}}^{2}}{M_{S}^{2}%
}\right)  ^{3/2}\quad;\quad\alpha\neq\beta.
\end{equation}
Because of the fact that the $f_{\alpha\beta}$ couplings are antisymmetric, there
are six decay channels in this class of models, whereas, experimentally, one
can observe only three distinct signals since neutrinos are indistinguishable,
i.e., charged lepton and missing energy [$Br (S^{\pm}\rightarrow\ell_{\alpha
}^{\pm}+\slashed E)=\sum_{\beta\neq\alpha}Br (S^{\pm}\rightarrow\ell_{\alpha
}^{\pm}\nu_{\beta})$].

\begin{figure}[h]
\begin{centering}
\includegraphics[width=0.5\textwidth]{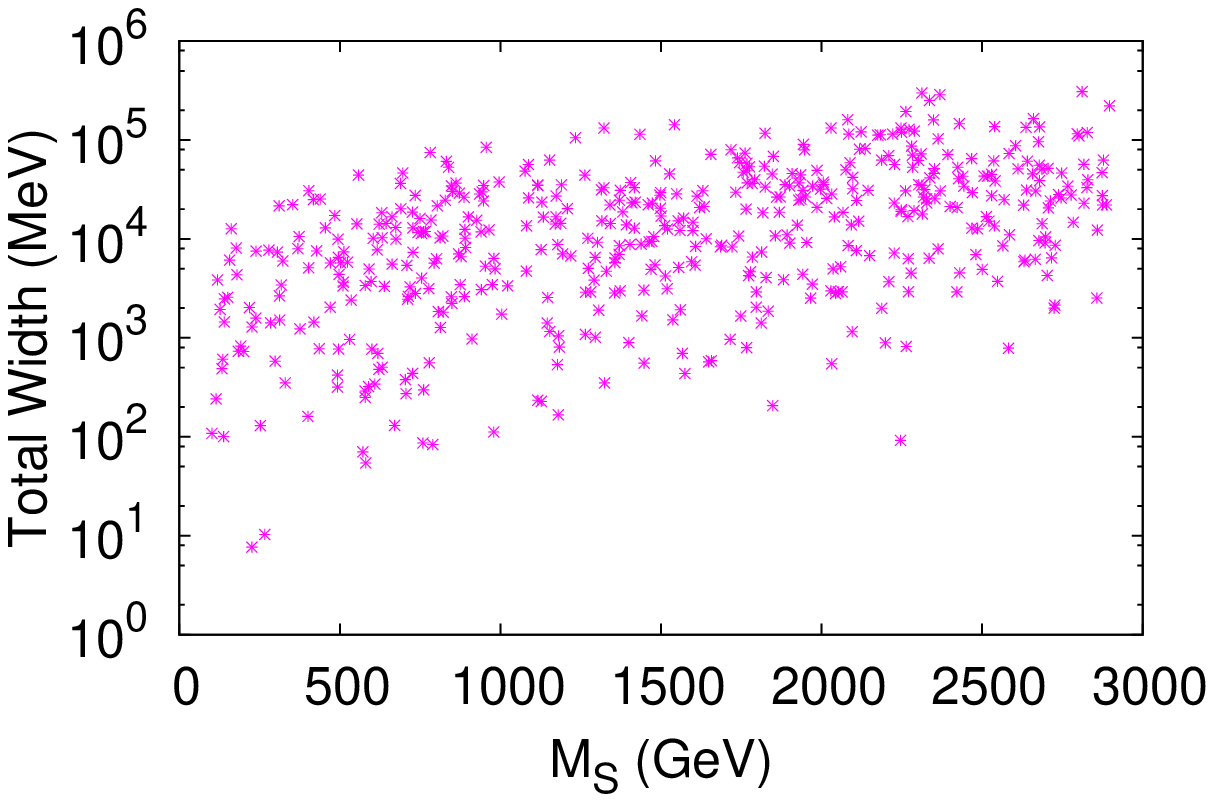}~\includegraphics[width = 0.5\textwidth]{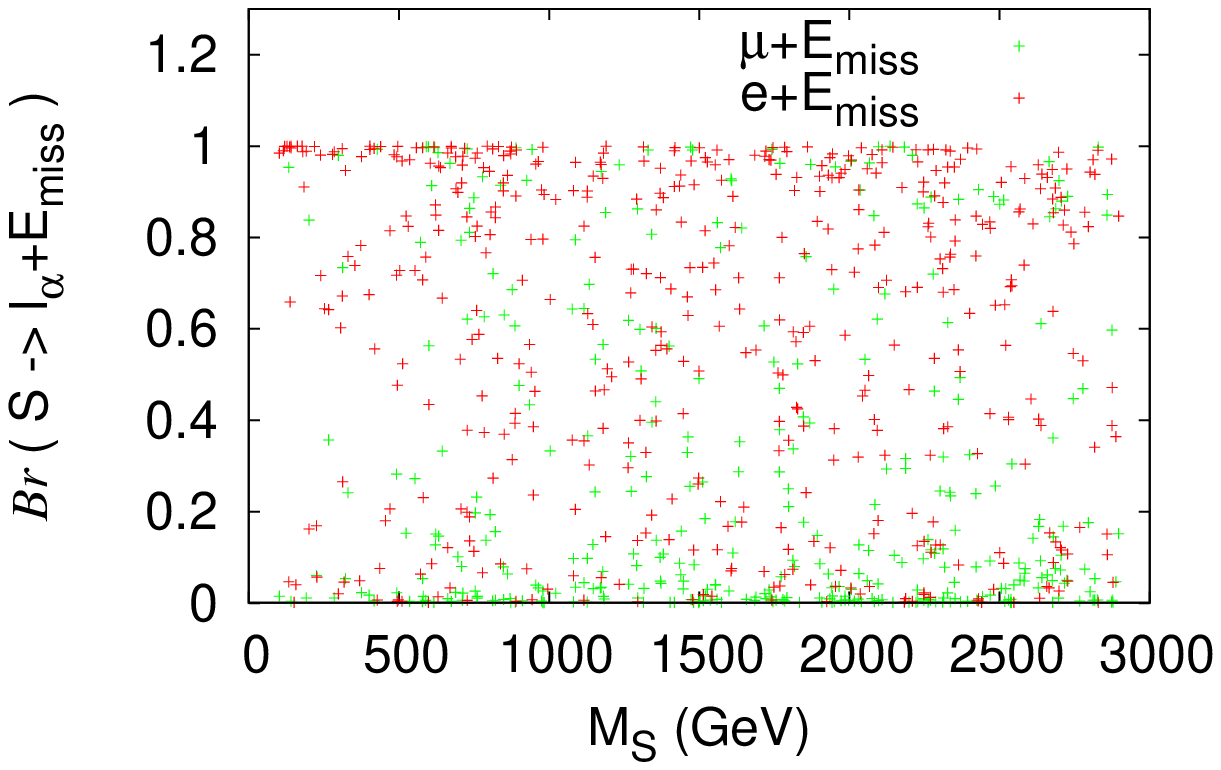}
\par\end{centering}
\caption{The charged scalar total decay width (left) and its different
branching ratios (right) vs $M_{S}$. The considered values of the charged
scalar mass and the Yukawa couplings $f_{\alpha\beta}$ are in agreement with
the experimental constraints mentioned above.}%
\label{knt-d}%
\end{figure}

For the same benchmark points used in Fig.~\ref{knt-p}, we show in
Fig.~\ref{knt-d} the charged scalar total decay width (left) and the obtained
different branching ratios $Br (S^{-}\rightarrow\ell_{\alpha}^{-}+\slashed E)$
(right) as a function of $M_{S}$.

Furthermore, as can be seen from Fig.~\ref{knt-d}, the total decay width of
the charged scalar is increasing for masses, and the branching ratios are
comparable for some benchmark points. In Sec. III, we choose a benchmark
point and investigate in detail the possibility of having a significant excess
in the process $S^{\pm}\rightarrow\ell_{\alpha}^{\pm}+\slashed E$. Such excess
will be a direct proof of the effect of the charged scalars.

\section{Signature of the charged scalar at the LHC}

In this section, we discuss one of the important tests of this
%a possibility of testing this
class of neutrino mass models (\ref{L}) which can be explored at the LHC TeV
energies. Hereafter, we consider one benchmark with the following parameters
values%
\begin{equation}
\left\{
\begin{array}
[c]{l}%
f_{e\mu}=-(4.97+i1.41)\times10^{-2},\quad f_{e\tau}=0.106+i0.0859,\\
f_{\mu\tau}=(3.04-i4.72)\times10^{-6},\quad\quad M_{S}=914.2\text{
\textrm{GeV.}}%
\end{array}
\right.  \label{fm}%
\end{equation}

These considered parameters values are in agreement with LFV bounds and could
be part of a benchmark point in such a model in which the neutrino oscillation data
and DM relic density can be easily addressed. The signature of a charged
scalar at the LHC will be identified through the detection of two opposite-sign charged leptons plus missing energy corresponding to the SM neutrinos.
Therefore, the final state consists of requiring dileptons plus missing
energy which will define the event topology of the signal,
\begin{equation}
pp\rightarrow\ell_{\alpha}^{\pm}\ell_{\beta}^{\mp}+\slashed E, \label{pp}%
\end{equation}
where $\ell_{\alpha}^{\pm}\ell_{\beta}^{\mp}$ can be $e^{+}$$e^{-}$, $e^{-}%
$$\mu^{+}$, or $\mu^{-}$ $\mu^{+}$, except the $\tau$ leptons which are very
difficult to identify experimentally. The missing energy $\slashed E$
contribution corresponds to any of the two neutrino final states
$\{\nu_{\alpha}\overline{\nu}_{\beta}\}$ with $\ell_{\alpha}\ell_{\beta}%
=e,\mu,\tau$ which also might include the background events produced from the
leptonic decay of $W^{\pm}$ gauge bosons.

Searches for a typical signal including new charged scalars $S^{\pm}$ through
kinematic distributions in events with dilepton final states have not yet found yet
any significant deviation from the SM as has been illustrated by ATLAS
\cite{atlas-c} and CMS~\cite{CMS-c} collaborations. The dominant source of
background is defined as being any process where $WW$, $ZZ$, or $Z\gamma$ are
intermediate states:
\begin{align}
pp  &  \rightarrow W^{+}W^{-}\rightarrow\ell_{\alpha}^{+}\ell_{\beta}^{-}%
\nu_{\alpha}\bar{\nu}_{\beta},\nonumber\\
pp  &  \rightarrow ZZ\left(  \gamma Z\right)  \rightarrow\ell_{\alpha}^{+}%
\ell_{\alpha}^{-}\nu_{\beta}\bar{\nu}_{\beta}. \label{BG}%
\end{align}

According to the interactions in Eq. (\ref{L}), the same final state can be
achieved through intermediate states that include a single or a pair of
charged scalars. Thereby, in this work, we are looking for any deviation from
the SM where the effective cross section in question is the difference between
the cross section estimated according to Eq. (\ref{L}) and the one estimated
within the SM. Thus, when finding the different cuts, we consider the values
of kinematic variables for which the effective differential cross section is
strictly positive.

In this work, the model files were produced using the LanHEP
program~\cite{Lanhep} for the Feynman rules generation in momentum
representation. Then the event generation and simulation for the
corresponding cross sections of both signal and background processes
at various c.m. energies were obtained by using
CalcHEP~\cite{Calchep}.

\begin{figure}[h]
\begin{centering}
\includegraphics[width = 0.5\textwidth]{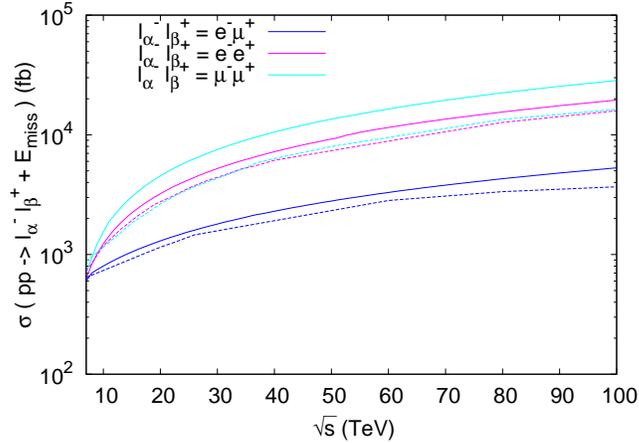}
\par\end{centering}
\caption{The cross section (in fb units) of different contributions (solid
lines) of the missing energy for each process as a function of the center of
mass (in TeV). Here, the dashed lines represent the SM values.}%
\label{pcm}%
\end{figure}

To probe the effect of lepton universality violating processes due to
the interactions given in Eq. (\ref{L}), we show in Fig.~\ref{pcm} the cross
section $\sigma(pp\rightarrow\ell_{\alpha}^{\pm}\ell_{\beta}^{\mp}+\slashed
E)$ that corresponds to the three cases ($\ell_{\alpha}\ell_{\beta}=ee,e\mu
,\mu\mu$) vs the c.m. energies $\sqrt{s}=[7~\text{TeV}-100~\text{TeV}]$.

As can be seen, for the SM case, the two processes $pp\rightarrow e^{+}%
e^{-}+\slashed E$ and $pp\rightarrow\mu^{+}\mu^{-}+\slashed E$ have the same
cross section values, while in our model it shows an increasing difference
with respect to the c.m. energy values. This can be understood due to the fact
that the charged scalar couplings to the leptons are not universal.
Therefore, it is possible to test these interactions when more data are
accumulated at the LHC Run II.

\subsection{Signal and background}

The criteria used to reject the background are based on the full kinematic
study of the events. For that, we have examined the kinematic distributions of
the signal and background to define a convenient set of cuts that can provide a
good discrimination against the reducible background.

To optimize the signal significance, the event selection is performed
in two steps: an initial \textit{preselection} and \textit{final selection}.
In the preselection we use an accurate cut on the $M_{T2}$ event variable\cite{mt2, mt22} to separate as much as possible the signal from the SM
background without a large reduction in the signal efficiency. This is applied
by imposing an inclusive cut on $M_{T2}$ defined as
\begin{equation}
M_{T2}=\underset{\vec{p}_{T}^{(\nu_{\alpha})}+\vec{p}_{T}^{(\bar{\nu}_{\beta
})}=\vec{p}_{T}^{miss}}{\min}\left[  \max\left(  M_{T}^{(\alpha)}%
+M_{T}^{(\beta)}\right)  \right]  , \label{mt2var}%
\end{equation}
with%
\begin{equation}
M_{T}^{(\alpha)}=m_{\ell_{\alpha}}^{2}+2\left(  E_{T}^{\ell_{\alpha}}%
E_{T}^{\nu_{\alpha}}-\vec{p}_{T}^{\ell_{\alpha}}.\vec{p}_{T}^{\nu_{\alpha}%
}\right)  ,
\end{equation}
where $E_{T}^{\ell_{\alpha}}$ ($E_{T}^{\nu_{\alpha}}$) and $\vec{p}_{T}%
^{\ell_{\alpha}}$ ($\vec{p}_{T}^{\nu_{\alpha}}$) are the energy and transverse
momentum energy of the charged lepton (neutrino). The $M_{T2}$ variable, which
is the transverse momentum imbalance in an event, is used to bound the masses
and to reflect the masses of the pair-produced particles. In the limit of
massless missing energy particles, the $M_{T2}$ variable can be written as
\cite{Chatrchyan:2012jx}
\begin{equation}
M_{T2}^{2}=2p_{T}^{\ell_{\alpha}}p_{T}^{\ell_{\beta}}\left(  1+\cos
\theta_{\alpha\beta}\right)  ,
\end{equation}
and $\theta_{\alpha\beta}\equiv\left(  \ell_{\alpha},\ell_{\beta}\right)  $ is
the angle between the charged leptons. As a result, in pair-production events
with two leptons and missing energy, the $M_{T2}$ distribution has an upper
limit at the mass of the mother particle. Therefore, the background events of
the charged leptons and neutrinos coming from the $W^{+}W^{-}$ source can be
eliminated by imposing an inclusive-$M_{T2}$ cut expressed as
\begin{equation}
M_{T2}>M_{W}. \label{MT2}%
\end{equation}

After imposing the condition (\ref{MT2}), we study different
distributions of the considered processes at both 8 and 14 TeV c.m.
energies , and then deduce the relevant kinematic cuts as summarized
in Table ~\ref{Tab}.

\begin{table}[h]
\centering
\begin{tabular}
[c]{cclc}\hline
Process & Cuts@8 TeV &  & Cuts@14 TeV\\\hline\hline
$p$$p$ $\rightarrow$ $e^{-}$$\mu^{+}+\slashed E$ & $%
\begin{array}
[c]{cc}%
80<p_{T}^{e^{-}}<250 & 80<p_{T}^{\mu^{+}}<270\\
-1.560<\eta_{e^{-}}<2.99 & -1.92<\eta_{\mu^{+}}<3
\end{array}
$ &  & $%
\begin{array}
[c]{cc}%
p_{T}^{e^{-}}>180 & p_{T}^{\mu^{+}}>170\\
1.1<\eta_{e^{-}}<2.89 & 1.2<\eta_{\mu^{+}}<3.02
\end{array}
$\\\hline\hline
$p$$p$ $\rightarrow$ $e^{-}$$e^{+}+\slashed E$ & $%
\begin{array}
[c]{c}%
25<p_{T}^{l}<120\\
-2.09<\eta_{l}<2.89
\end{array}
$ &  & $%
\begin{array}
[c]{c}%
30<p_{T}^{l}<80\\
-2.8<\eta_{l}<2.95
\end{array}
$\\\hline\hline
$p$$p$ $\rightarrow$ $\mu^{-}$$\mu^{+}+\slashed E$ & $%
\begin{array}
[c]{c}%
30<p_{T}^{l}<155\\
-2.38<\eta_{l}<2.1
\end{array}
$ &  & $%
\begin{array}
[c]{c}%
25<p_{T}^{l}<40\\
-0.13<\eta_{l}<3
\end{array}
$\\\hline
\end{tabular}
\caption{The considered cuts for the three final states at
$\sqrt{s}$= 8 and 14 TeV center-of-mass energy. The $p_{T}^{\ell}$
and $\eta_{\ell} $ are, respectively, the transverse momentum and
pseudorapidity of the charged lepton ($e,\mu$).} \label{Tab}
\end{table}

Note that some cuts are set as upper bounds on charged leptons energies and
transverse momenta. These are required to avoid a negative effective deferential
cross section. When comparing the charged leptons cut values of the two
processes $pp\rightarrow e^{+}e^{-}+\slashed E$ and $pp\rightarrow\mu^{+}%
\mu^{-}+\slashed E$, one remarks that an $e-\mu$ asymmetry exists, which gets
larger at higher c.m. energies. This can be understood due to nonuniversal
couplings of the charged scalars to a lepton in Eq. (\ref{L}) as it has been stated
in the previous section.

\subsection{Numerical results}

The event yields which pass the kinematic cuts under the conditions
described in Table ~\ref{Tab} will be used to evaluate the physics
significance of the signal. We followed a detailed kinematic
variables scan where a tightened cut selection is taken into account
in order to optimize the separation of the signal to the background
ratio. For each signal final state, the signal significance is
defined as
\begin{equation}
S=\frac{N_{ex}}{\sqrt{N_{ex}+N_{B}}},
\end{equation}
where $N_{ex}$ is the excess events number of the considered signal and
$N_{B}$ is the number of events of the background contributions that can mimic
the signal. More explicitly, this excess can be expressed as
\begin{equation}
N_{ex}=N_{M}-N_{B}=L\times(\sigma^{M}-\sigma^{B}),
\end{equation}
where $N_{M}$ is the expected events number due to all the new model
interactions including those of the SM, $L$ is the integrated
luminosity (known also as $\int\mathcal{L}dt$), and $\sigma^{M}$
($\sigma^{B}$) is the total expected (background) cross section. In
summary, Table ~\ref{tab-S} presents the signal and background cross
sections at 8 and 14 \textrm{TeV} for the three
different signatures after passing the selection cuts given earlier.
The corresponding significance depending on the available (and the
expected with 14 \textrm{TeV}) luminosity is then extracted and
written in the last column of each table.

\begin{table}[h]
\centering%
\begin{tabular}
[c]{|c|cccc|c|cccc|}\hline
&  &  & $\sqrt{s}=8~TeV$ &  &  &  &  & $\sqrt{s}=14~TeV$ & \\\cline{2-5}%
\cline{7-10}%
Process & $\sigma^{M}(fb)$ & \multicolumn{1}{|c}{$\sigma^{B}(fb)$} &
\multicolumn{1}{|c}{$(\sigma^{M}-\sigma^{B})/\sigma^{B}$} &
\multicolumn{1}{|c|}{$S_{20}$} &  & $\sigma^{M}(fb)$ &
\multicolumn{1}{|c}{$\sigma^{B}(fb)$} & \multicolumn{1}{|c}{$(\sigma
^{M}-\sigma^{B})/\sigma^{B}$} & \multicolumn{1}{|c|}{$S_{100}$}\\\hline\hline
$pp\rightarrow e^{-}\mu^{+}+\slashed E$ & 13.03 & \multicolumn{1}{|c}{11.98} &
\multicolumn{1}{|c}{0.0876} & \multicolumn{1}{|c|}{1.301} &  & 1.253 &
\multicolumn{1}{|c}{0.459} & \multicolumn{1}{|c|}{1.7} &
\multicolumn{1}{|c|}{7.093}\\\cline{1-5}\cline{7-10}%
$pp\rightarrow e^{-}e^{+}+\slashed E$ & 62.74 & \multicolumn{1}{|c}{59.72} &
\multicolumn{1}{|c}{0.0506} & \multicolumn{1}{|c|}{1.7051} &  & 44.45 &
\multicolumn{1}{|c}{38.65} & \multicolumn{1}{|c|}{0.150} &
\multicolumn{1}{|c|}{8.699}\\\cline{1-5}\cline{7-10}%
$pp\rightarrow\mu^{-}\mu^{+}+\slashed E$ & 81.691 & \multicolumn{1}{|c}{77.49}
& \multicolumn{1}{|c}{0.0542} & \multicolumn{1}{|c|}{2.0786} &  & 65.27 &
\multicolumn{1}{|c}{56.86} & \multicolumn{1}{|c|}{0.148} &
\multicolumn{1}{|c|}{10.409}\\\hline
\end{tabular}
\caption{The cross section of the total expected signals
($\sigma^{M} $) and the corresponding background ($\sigma^{B}$) are
used to estimate the significance $S_{20}$ ($S_{100}$)\ at 8
\textrm{TeV} (14 \textrm{TeV}) with the recorded integrated
luminosity $L=20~fb^{-1}$ ($L=100~fb^{-1}$).}
\label{tab-S}%
\end{table}

\begin{figure}[h]
\begin{center}
\includegraphics[width=0.5\textwidth]{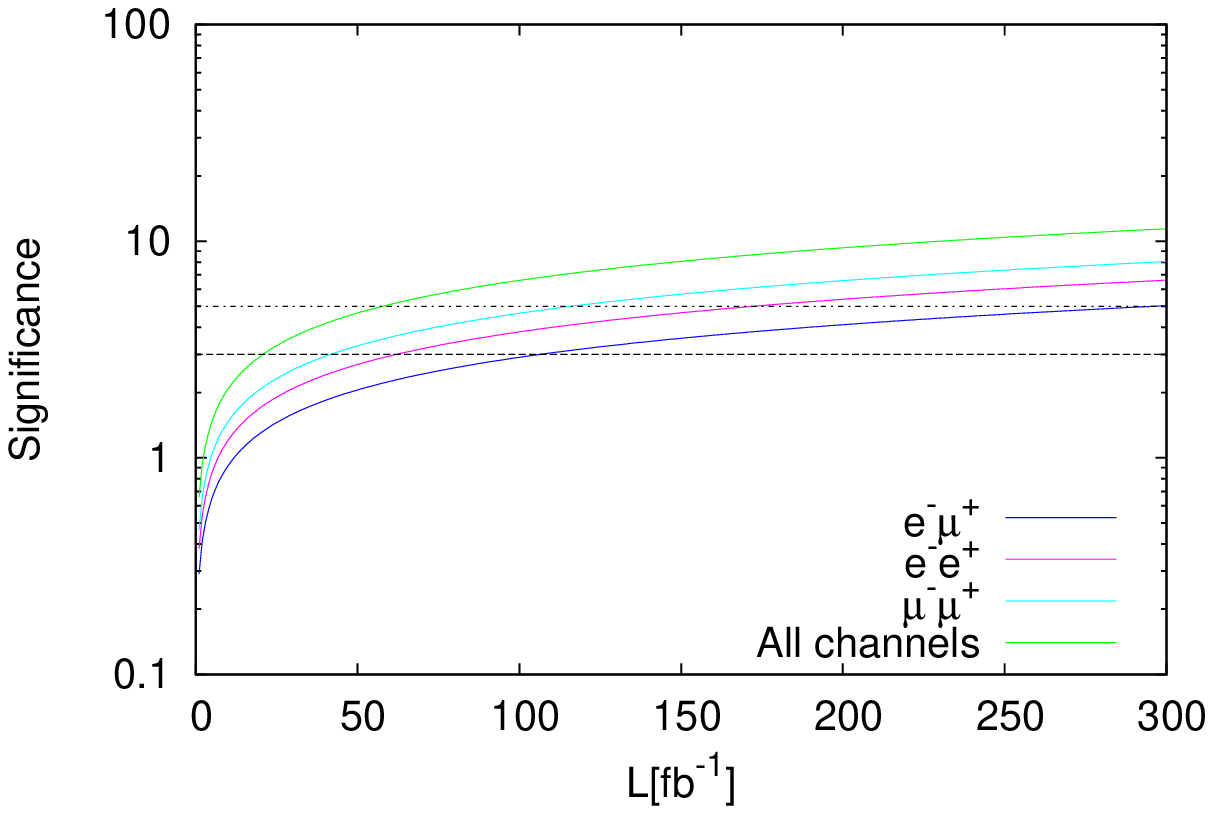}~\includegraphics[width=0.5\textwidth]{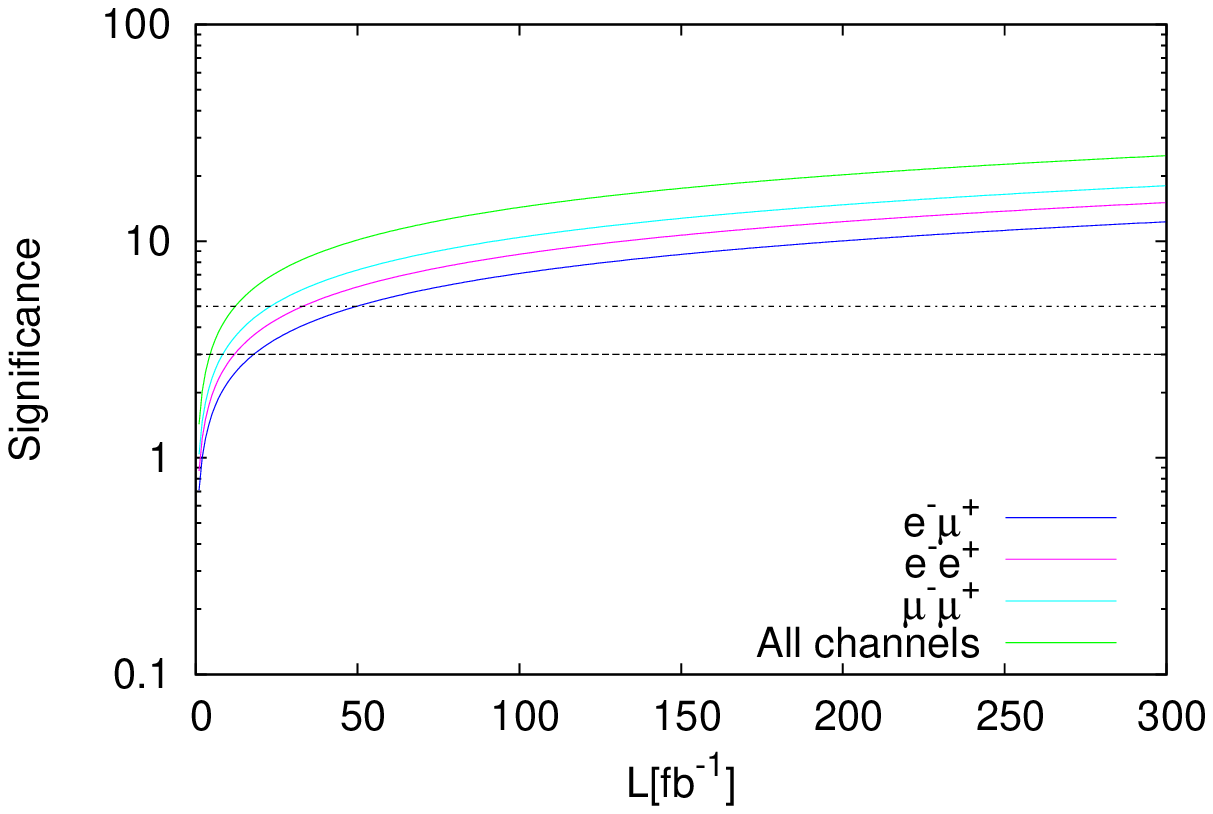}
\end{center}
\caption{The significance vs the luminosity at $\sqrt{s}=8~TeV$ (left) and
at $\sqrt{s}=14~TeV$ (right) for the three different signatures. The two
horizontal dashed lines in each panel indicate the corresponding significance
values for $S=3$ and $S=5$, respectively.}%
\label{l}%
\end{figure}

The numerical results show that the signal significance varies in the range of
[1.30 - 2.07] at $\sqrt{s}=$ 8 \textrm{TeV} with an optimum value up to 10.41
at $\sqrt{s}=$ 14 \textrm{TeV}. It should be pointed out that the final states
with the charged leptons $\mu^{-}\mu^{+}$ are significantly larger than
$e^{-}e^{+}$, and as a consequence, it is the most favorable channel to tackle
the singlet charged pair production at both collider energies. A convenient
representation of these results is also illustrated in Fig.~\ref{l} where the
signal significance is plotted against the available and expected luminosity regions.

\begin{figure}[h]
\begin{center}
\includegraphics[width=0.5\textwidth]{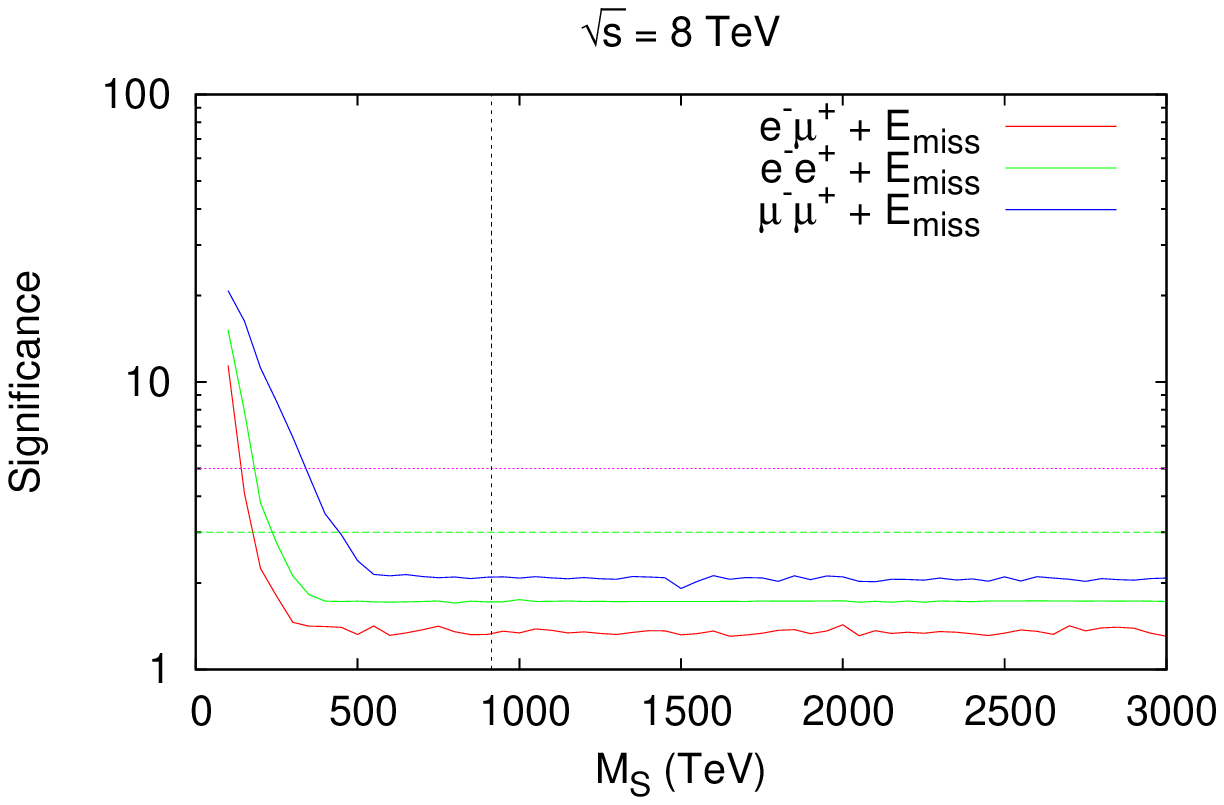}~\includegraphics[width=0.5\textwidth]{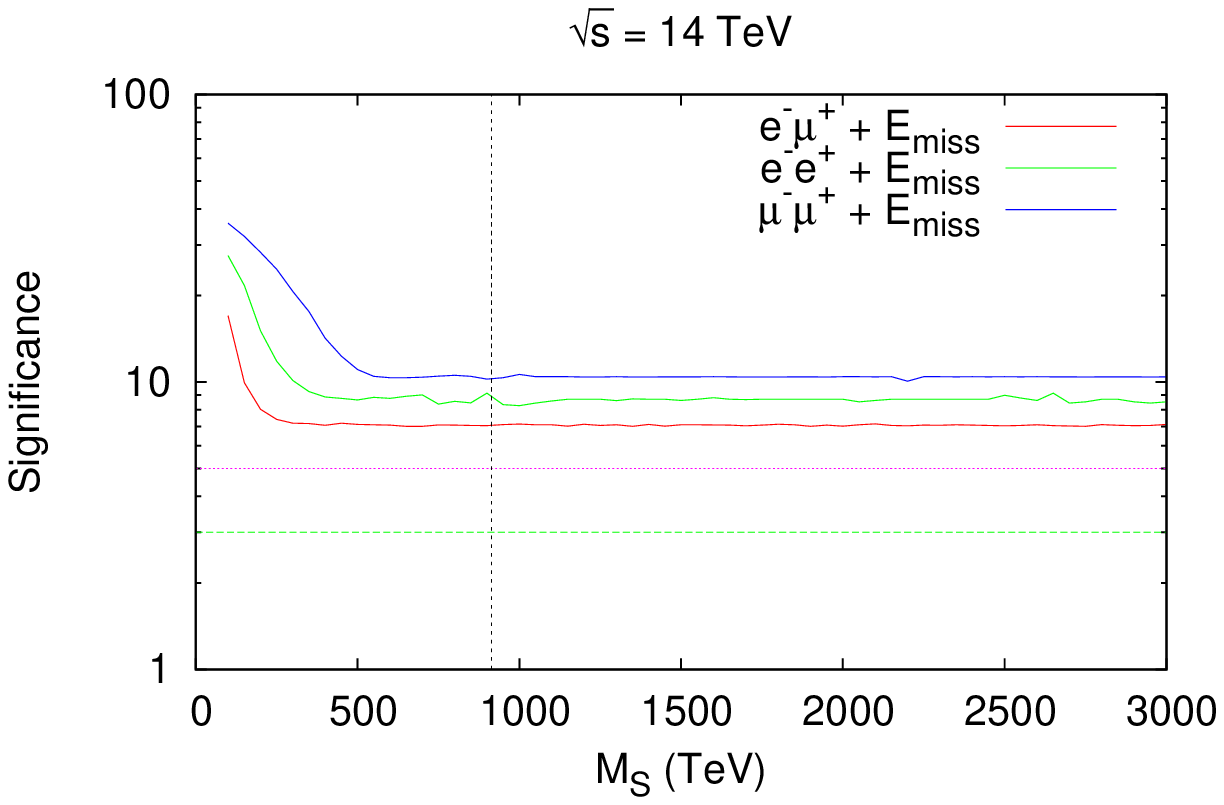}
\end{center}
\caption{The significance vs the mass of the charged scalar
$M_{S}$ for each channel at L = 20 $fb^{-1}$ (left) and at L = 100
$fb^{-1}$ (right). The two dashed horizontal lines indicate the
significance values at S = 3 and S = 5,   and the
vertical one indicates the charged scalar mass given (\ref{fm}).}
\label{ms1}
\end{figure}

Having shown the consistency of all the possible signal signatures
at both energies, it is justified to explore furthermore the phase
space to determine their behavior based on one of the main model
parameters which is the mass of the charged scalar. In
Fig.~\ref{ms1}, we show the dependence of the significance vs
$M_{S}$ at both c.m. energies 8 and 14 TeV with
$L=20~fb^{-1}$ (left) and  $L=100~fb^{-1}$ (right). In this
figure, the branching fractions for the LFV processes $\ell_{\alpha}\rightarrow\ell_{\beta}+\gamma$ have been taken  to
be equal to the value computed using  the parameters given in Eq.
(\ref{fm}) and which are just below the LFV bounds.

As can be seen in Fig.~\ref{ms1}, left, the LHC Run I data can be used to
exclude the charged scalars with masses below 400 GeV, whereas Fig.~\ref{ms1}%
, right, shows that the signal is likely to be seen at LHC@14. Note that for
benchmarks with $Br(\ell_{\alpha}\rightarrow\ell_{\beta}+\gamma)$ much smaller
than the current experimental bound the significance gets reduced since in
this case the couplings $f_{\alpha\beta}$ are tiny and/or $M_{S}$ is very large.

\section{Conclusion}

In this work, we have investigated the possibility of testing the singlet
charged scalars effect in a class of neutrino mass models. This dedicated
search at the LHC energies can place stringent limits on the cross section
times the branching ratios of the charged scalar decaying to leptons in the
radiative neutrino mass models. We considered the opposite-sign dilepton final
states with missing energy at the LHC proton-proton collisions with both 8 and
14 TeV c.m. energies. An observed deviation from the SM, if seen in the future,
could be a very important hint that leads us to consider the SM left-handed
neutrinos as Majorana fermions. We found that these charged scalars can be
pair produced at the LHC and decay inside the detector. To extract the
contribution of the charged scalars, we analyzed the dilepton final states
with missing energy ($pp\rightarrow\ell_{\alpha}^{\pm}\ell_{\beta}^{\mp
}+\slashed E$), where $\ell_{\alpha}\ell_{\beta}=ee,e\mu,\mu\mu$. We concluded
that an inclusive cut on the $M_{T2}$\ event variable is vital and has an
effective suppression of the large SM background. We found that at the LHC@8
TeV the charged scalars effect on $pp\rightarrow\ell_{\alpha}^{\pm}%
\ell_{\beta}^{\mp}+\slashed
E$ cannot be seen with the integrated luminosity $L=20~fb^{-1}$, and no
significant deviations are observed. However, at the LHC@14 TeV with the
expected integrated luminosity $L=100~fb^{-1}$, an effect can be found in all
channels. Moreover, it has been shown that the signal significance decreases
rapidly with the increasing charged scalar mass. For instance, when
$M_{S}=912.4$ GeV, it leads to 3$\sigma$ significance of for $L=20$ $fb^{-1}$
and more than 5$\sigma$ significance for $L=100$ $fb^{-1}$, and the favored
channel would be $pp\rightarrow\mu^{+}\mu^{-}+\slashed E$ at both 8, 14 TeV c.m. energies.

\subsection*{Acknowledgments}

We would like to thank K. Cheung for his useful comments and A.B. Hammou for
reading the manuscript. A. A., C. G., and D. C. would like to thank the ICTP for the
warm hospitality during part of this work. A. A. is supported by the Algerian
Ministry of Higher Education and Scientific Research under the CNEPRU Project
No. D01720130042.


\begin{thebibliography}{99}                                                                                               %


\bibitem {Zee-Ma}A. Zee, Phys. Lett. B \textbf{161}, 141 (1985); E. Ma, Phys. Rev.
Lett. \textbf{81}, 1171 (1998).

\bibitem {Zee-86}A. Zee, Nucl. Phys. \textbf{B 264}, 99 (1986); K. S. Babu, Phys. Lett.
B \textbf{203}, 132 (1988); M.~Aoki, S.~Kanemura, T.~Shindou, and K.~Yagyu,
%``An R-parity conserving radiative neutrino mass model without right-handed neutrinos,''
J. High Energy Phys. 10 \textbf{1007} (2010) 084; J. High Energy Phys. 11 (2010) 049; G.~Guo, X.~G.~He, and G.~N.~Li,
%``Radiative Two Loop Inverse Seesaw and Dark Matter,''
J. High Energy Phys. 10 (2012) 044; Y.~Kajiyama,
H.~Okada, and K.~Yagyu,
%``Two Loop Radiative Seesaw Model with Inert Triplet Scalar Field,''
Nucl.\ Phys.\ \textbf{B874}, 198 (2013).

\bibitem {Aoki:2008av}M.~Aoki, S.~Kanemura, and O.~Seto,
%``Neutrino mass, Dark Matter and Baryon Asymmetry via TeV-Scale Physics without Fine-Tuning,''
Phys.\ Rev.\ Lett.\ \textbf{102}, 051805 (2009);
%%CITATION = doi:10.1103/PhysRevLett.102.051805;%%
M.~Aoki, S.~Kanemura, and O.~Seto,
%``A Model of TeV Scale Physics for Neutrino Mass, Dark Matter and Baryon Asymmetry and its Phenomenology,''
Phys.\ Rev.\ D \textbf{80}, 033007 (2009).

\bibitem {knt}L.~M.~Krauss, S.~Nasri, and M.~Trodden,
%``A Model for neutrino masses and dark matter,''
Phys.\ Rev.\ D \textbf{67}, 085002 (2003).
%%CITATION = doi:10.1103/PhysRevD.67.085002;%%


\bibitem {knt3}A.~Ahriche, C.~S.~Chen, K.~L.~McDonald, and S.~Nasri,
%``Three-loop model of neutrino mass with dark matter,''
Phys.\ Rev.\ D \textbf{90}, 015024 (2014).
%%CITATION = doi:10.1103/PhysRevD.90.015024;%%


\bibitem {knt5}A.~Ahriche, K.~L.~McDonald, and S.~Nasri,
%``A Model of Radiative Neutrino Mass: with or without Dark Matter,''
J. High Energy Phys. 10 (2014) 167.
%%CITATION = doi:10.1007/JHEP10(2014)167;%%


\bibitem {knt7}A.~Ahriche, K.~L.~McDonald, S.~Nasri, and T.~Toma,
%``A Model of Neutrino Mass and Dark Matter with an Accidental Symmetry,''
Phys.\ Lett.\ B \textbf{746}, 430 (2015).
%%CITATION = doi:10.1016/j.physletb.2015.05.031;%%


\bibitem {Ahriche:2015loa}A.~Ahriche, K.~L.~McDonald, and S.~Nasri,
%``A Radiative Model for the Weak Scale and Neutrino Mass via Dark Matter,''
  JHEP {\bf 1602} (2016) 038
%  doi:10.1007/JHEP02(2016)038
  [arXiv:1508.02607 [hep-ph]].


\bibitem {hna}A.~Ahriche, and S.~Nasri,
%``Dark matter and strong electroweak phase transition in a radiative neutrino mass model,''
J. Cosmol. Astropart. Phys. 07 (2013) 035.
%%CITATION = doi:10.1088/1475-7516/2013/07/035;%%


\bibitem {ILC}A.~Ahriche, S.~Nasri, and R.~Soualah,
%``Radiative Neutrino Mass Model at the $e^{-}e^{+}$ Linear Collider,''
Phys.\ Rev.\ D \textbf{89}, 095010 (2014).

\bibitem {AMN}A.~Ahriche, K.~L.~McDonald, and S.~Nasri,
%``Scalar Sector Phenomenology of Three-Loop Radiative Neutrino Mass Models,''
Phys.\ Rev.\ D \textbf{92}, 095020 (2015).

\bibitem {osland}P.~Osland, A.~Pukhov, G.~M.~Pruna, and M.~Purmohammadi,
%``Phenomenology of charged scalars in the CP-Violating Inert-Doublet Model,''
J. High Energy Physics. 04 (2013) 040.
%%CITATION = doi:10.1007/JHEP04(2013)040;%%


\bibitem {Cheung}K.~Cheung and O.~Seto,
%``Phenomenology of charged scalars in the CP-Violating Inert-Doublet Model,''
Phys. Rev. D \textbf{69}, 113009 (2004).
%%CITATION = doi:10.1007/JHEP04(2013)040;%%


\bibitem {FLV-bound}F.~del \'{A}guila and M.~Chala,
%``LHC bounds on Lepton Number Violation mediated by doubly and singly-charged scalars,''
J. High Energy Phys. 03 (2014) 027.
%%CITATION = doi:10.1007/JHEP03(2014)027;%%


\bibitem {seesaw}E.~J.~Chun, and P.~Sharma,
%``Search for a doubly-charged boson in four lepton final states in type II seesaw,''
Phys.\ Lett.\ B \textbf{728}, 256 (2014).
%%CITATION = doi:10.1016/j.physletb.2013.11.056;%%


\bibitem {Adam:2013mnn}J.~Adam \textit{et al.} (MEG Collaboration),
%``New constraint on the existence of the $\mu^+ \to e^+\gamma$ decay,''
Phys.\ Rev.\ Lett.\ \textbf{110}, 201801 (2013).

\bibitem {ATLAS}G. Aad \textit{et al.} (ATLAS Collaboration), Phys. Lett. B \textbf{716}, 1 (2012).

\bibitem {CMS}S. Chatrchyan \textit{et al.} (CMS Collaboration), Phys. Lett. B \textbf{716}, 30 (2012).

\bibitem {PDG}K.~A.~Olive \textit{et al.} (Particle Data Group
Collaboration),
%``Review of Particle Physics,''
Chin.\ Phys.\ C \textbf{38}, 090001 (2014).

\bibitem {kanemura}S.~Kanemura, T.~Nabeshima, and H.~Sugiyama,
%``Radiative type-I seesaw model with dark matter via U(1)$_{B-L}$ gauge symmetry breaking at future linear colliders,''
Phys.\ Rev.\ D \textbf{87}, 015009 (2013).
%%CITATION = doi:10.1103/PhysRevD.87.015009;%%


\bibitem {atlas-c}G.~Aad \textit{et al.} (ATLAS Collaboration),
%``Search for doubly-charged Higgs bosons in like-sign dilepton final states at $\sqrt{s}=7$ TeV with the ATLAS detector,''
Eur.\ Phys.\ J.\ C \textbf{72}, 2244 (2012).
%%CITATION = doi:10.1140/epjc/s10052-012-2244-2;%%


\bibitem {CMS-c}S.~Chatrchyan \textit{et al.} (CMS Collaboration),
%``A search for a doubly-charged Higgs boson in $pp$ collisions at $\sqrt{s}=7$ TeV,''
Eur.\ Phys.\ J.\ C \textbf{72}, 2189 (2012).
%%CITATION = doi:10.1140/epjc/s10052-012-2189-5;%%


\bibitem {Lanhep}A.~Semenov,
%``LanHEP: A Package for the automatic generation of Feynman rules in field theory. Version 3.0,''
Comput.\ Phys.\ Commun.\ \textbf{180}, 431 (2009).
%%CITATION = doi:10.1016/j.cpc.2008.10.012;%%


\bibitem {Calchep}A.~Belyaev, N.~D.~Christensen, and A.~Pukhov,
%``CalcHEP 3.4 for collider physics within and beyond the Standard Model,''
Comput.\ Phys.\ Commun.\ \textbf{184}, 1729 (2013).
%%CITATION = doi:10.1016/j.cpc.2013.01.014;%%


\bibitem {mt2}C.~G.~Lester, and D.~J.~Summers,
%``Measuring masses of semiinvisibly decaying particles pair produced at hadron colliders,''
Phys.\ Lett.\ B \textbf{463}, 99 (1999).
%%CITATION = doi:10.1016/S0370-2693(99)00945-4;%%


\bibitem {mt22}A.~Barr, C.~Lester, and P.~Stephens,
%``m(T2): The Truth behind the glamour,''
J.\ Phys.\ G \textbf{29}, 2343 (2003).
%%CITATION = doi:10.1088/0954-3899/29/10/304;%%


\bibitem {Chatrchyan:2012jx}S.~Chatrchyan \textit{et al.} (CMS
Collaboration),
%``Search for supersymmetry in hadronic final states using MT2 in $pp$ collisions at $\sqrt{s} = 7$ TeV,''
J. High Energy Phys. 10 (2012) 018.
%%CITATION = doi:10.1007/JHEP10(2012)018;%%

\end{thebibliography}
\end{document}